\documentclass[aps,prd,twocolumn,superscriptaddress,tightenlines,nofootinbib]{revtex4-1}

\usepackage{amsfonts,amsmath,amsthm,amssymb}
\usepackage{mathrsfs}
\usepackage{bm}
\usepackage{color}
\usepackage{epsfig}
\usepackage{mathrsfs}

\usepackage[usenames,dvipsnames]{xcolor}
\definecolor{orange}{cmyk}{0,0.5,1,0}
\definecolor{rossoCP3}{cmyk}{0,.88,.77,.40}
\definecolor{graa}{rgb}{0.8,0.8,0.8}
\definecolor{blaa}{rgb}{0.2,0.2,0.6}

\newcommand{\PRE}[1]{{#1}}   
\newcommand{\met} {\not\!\! E_T}

\newcommand{\beq}{\begin{equation}}
\newcommand{\eeq}{\end{equation}}
\newcommand{\bea}{\begin{flushleft} \begin{eqnarray}}
\newcommand{\eea}{\end{eqnarray}\end{flushleft}}

\newcommand{\postscript}[2]{\setlength{\epsfxsize}{#2\hsize}
   \centerline{\epsfbox{#1}}}
\newcommand{\comment}[1]{}

\newcommand{\ci}[1]{}

\newcommand{\ba}{\begin{eqnarray}}
\newcommand{\ea}{\end{eqnarray}}
\newcommand{\be}{\begin{equation}}
\newcommand{\ee}{\end{equation}}
\newcommand{\bay}[1]{\left(\begin{array}{#1}}
\newcommand{\eay}{\end{array}\right)}

\def\met{\mbox{${\hbox{$E$\kern-0.6em\lower-.1ex\hbox{/}}}_T$}} 

\newcommand{\beqa}{\begin{eqnarray}}
\newcommand{\eeqa}{\end{eqnarray}}

\begin{document}

\title{\color{rossoCP3}{ IceCube neutrinos, decaying dark matter, and
    the Hubble constant
}}

\author{Luis A. Anchordoqui}
\affiliation{Department of Physics and Astronomy, Lehman College, City University of
  New York, NY 10468, USA
}

\affiliation{Department of Physics, Graduate Center, City University
  of New York, 365 Fifth Avenue, NY 10016, USA
}

\affiliation{Department of Astrophysics, American Museum of Natural History, Central Park West  79 St., NY 10024, USA}
\author {Vernon Barger}
\affiliation{Department of Physics, University of Wisconsin, Madison, WI 53706, USA}

\author{Haim \nolinebreak Goldberg}
\affiliation{Department of Physics,
Northeastern University, Boston, MA 02115, USA
}

\author{Xing \nolinebreak Huang}
\affiliation{Department of Physics, 
National Taiwan Normal University, Taipei, 116, Taiwan
}

\author{Danny Marfatia} 
\affiliation{Department of Physics and
  Astronomy, University of Hawaii, Honolulu, HI 96822, USA}

\author{Luiz H. M. da Silva}
\affiliation{Department of Physics,
University of Wisconsin-Milwaukee,
 Milwaukee, WI 53201, USA
}

\author{Thomas J. Weiler}
\affiliation{Department of Physics and Astronomy,
Vanderbilt University, Nashville TN 37235, USA
}

\date{June 2015}

\PRE{\vspace*{.15in}}

\begin{abstract}
  \noindent Cosmological parameters deduced from the \emph{Planck}
  measurements of anisotropies in the cosmic microwave background are
  at some tension with direct astronomical measurements of various
  parameters at low redshifts.  Very recently, it has been conjectured
  that this discrepancy can be reconciled if a certain fraction of
  dark matter is unstable and decays between recombination and the
  present epoch. Herein we show that if the superheavy relics have a
  branching into neutrinos ${\cal B}_{X \to \nu \bar \nu} \sim 3
  \times 10^{-9}$, then this scenario can also accommodate the recently
  discovered extraterrestrial flux of neutrinos, relaxing the tension
  between IceCube results and \emph{Fermi} LAT data.  The model is fully
  predictive and can be confronted with future IceCube data. We
  demonstrate that in 10 years of observation IceCube will be able to
  distinguish the mono-energetic signal from $X$ decay at the $3\sigma$
  level. In a few years of data taking with the upgraded
  \emph{IceCube-Gen2} enough statistics will be gathered to elucidate
  the dark matter--neutrino connection at the $5\sigma$ level.
\end{abstract}

\maketitle

We propose an explanation for the origin of IceCube
neutrinos~\cite{Aartsen:2013bka} assuming heavy decaying dark matter,
which is characterized by a lifetime and a comoving number density
that may resolve the tension between \emph{Planck} data and low
redshift astronomical measurements~\cite{Berezhiani:2015yta}.  The
ensuing discussion will be framed in the context of dissipative dark
matter. We consider a multicomponent hidden sector, with strong self
interactions, featuring a massless dark photon which mixes with the
ordinary photon and a dark $Z'$ which mixes with the $Z$. The dark
photon induces a one-loop photon wave function mixing term $\epsilon
F^{\mu \nu} F'_{\mu\nu}$~\cite{Goldberg:1986nk}. Early universe
cosmology and galactic structure considerations constrain the dark
photon mixing strength, $\epsilon \alt
10^{-9}$~\cite{Petraki:2014uza}. Bounds on the $Z-Z'$ mixing angle are
less restrictive~\cite{Chun:2010ve}.

{\color{rossoCP3} {\bf \emph{Cosmic Neutrinos---}}} Recently, the
IceCube Collaboration reported the discovery of extraterrestrial
neutrinos~\cite{Aartsen:2013bka}. By establishing a strict veto
protocol, the collaboration was able to isolate 36 events in 3 years
of data, with energies between $30~{\rm TeV} \alt E_\nu \alt 2~{\rm
  PeV}$. These events follow the expected spectral shape ($\propto
E_\nu^{-2}$) of a Fermi engine, and are consistent with an isotropic
distribution in the sky. A purely atmospheric explanation of the data
can be excluded at 5.7$\sigma$.

At $E_\nu^{\rm res} \simeq 6.3~{\rm PeV}$, one expects to observe a
dramatic increase in the event rate for $\bar \nu_e$ in ice due to the
“Glashow resonance” in which $\bar \nu_e e^- \to W^- \to {\rm shower}$
greatly increases the interaction cross
section~\cite{Glashow:1960zz}. The hypothesis of an unbroken power law
$\propto E_\nu^{-\alpha}$ then requires $\alpha \agt 2.45$ to be
consistent with data at 1$\sigma$~\cite{Barger:2014iua}. More
recently, the IceCube search technique was refined to extend the
neutrino sensitivity to lower energies $E_\nu \agt 10~{\rm
  TeV}$~\cite{Aartsen:2014muf}.  A fit to the resulting data, assuming
a single unbroken power law and equal neutrino fluxes of all flavors,
finds a softer spectrum
\begin{eqnarray}
\Phi_{\rm IceCube}^{\rm per \, flavor} (E_\nu) & = & 2.06^{+0.4}_{-0.3} \times 10^{-18}
\left(\frac{E_\nu}{10^5~{\rm GeV}} \right)^{-2.46 \pm 0.12} \nonumber
\\
& \times & {\rm GeV}^{-1}
   \, {\rm cm}^{-2} \, {\rm sr}^{-1} \, {\rm s}^{-1}
\label{icecubeflux}
\end{eqnarray}
and already mildly excludes the benchmark spectral index $\alpha = 2$.

\begin{figure}[tbp]
\postscript{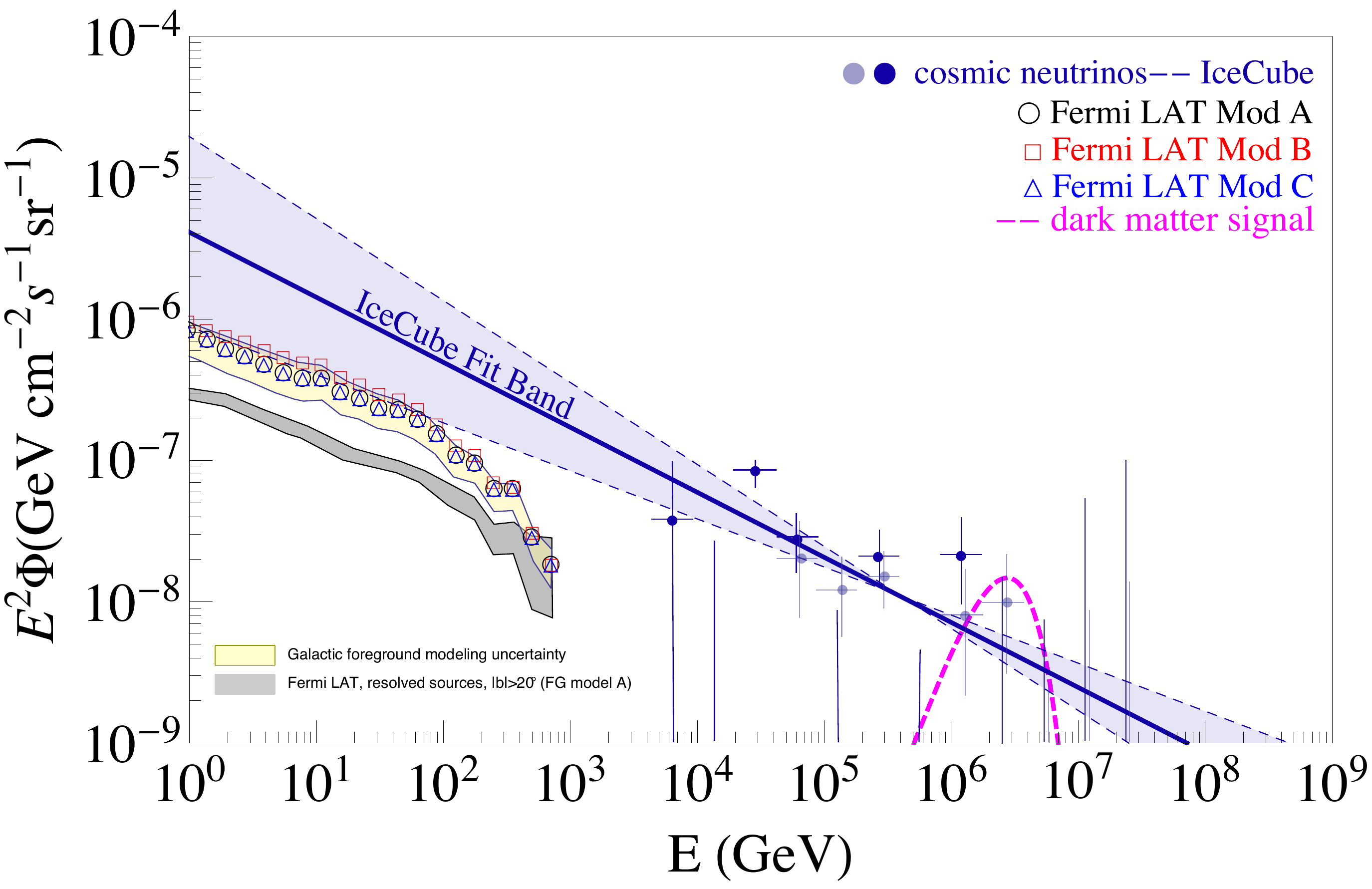}{0.9}
\caption{The open symbols represent the total extragalactic
  $\gamma$-ray background for different foreground (FG) models as
  reported by the Fermi LAT Collaboration~\cite{Ackermann:2014usa}. For details on the modeling of 
the diffuse Galactic foreground emission in the benchmark FG models A,
B and C, see~\cite{Ackermann:2014usa}. The cumulative intensity from
  resolved \emph{Fermi} LAT sources at latitudes $|b| > 20^\circ$ is indicated by a (grey)
  band. The solid symbols indicate the neutrino flux (per flavor) reported by
  the IceCube Collaboration. The blue points are from the data sample
  of the most recent IceCube analysis~\cite{Aartsen:2014muf}. The
  light grey data points are from the 3-year data sample
  of~\cite{Aartsen:2013bka}, 
shifted slightly to the right for better visibility.
The best fit to the data (extrapolated
  down to lower energies), is also shown
  for comparison~\cite{Aartsen:2014muf}. The dashed line indicates the mono-energetic signal
  from dark matter decay. Note that a plotting of $E^2 \Phi  = E
  dF/(d\Omega \, dA \, dt \, d \ln E )$ versus
  $\ln E$ conserves the area under a spectrum even after processing
  the electromagnetic cascade.
Thus, the area of the $\pi^0$ contribution to the diffuse $\gamma$-ray
spectrum (total diffuse $\gamma$-ray flux provides an upper bound)
implies the low energy cutoff (upper bound) to the $\pi^\pm$ origin of the neutrinos.
\label{fig:fermi+icecube+dm}}
\end{figure}

The neutrino flux in (\ref{icecubeflux}) is exceptionally high by astronomical standards, with a
magnitude comparable to the Waxman-Bahcall bound~\cite{Waxman:1998yy}.
A saturation of this bound can only be achieved within astrophysical
environments where accelerator and target are essentially integrated.
Potential candidate sources are discussed
in~\cite{Anchordoqui:2013dnh}. These powerful sources produce roughly
equal numbers of $\pi^0$, $\pi^+$ and $\pi^-$ in the proton-proton
beam dump. The $\pi^0$ accompanying the $\pi^\pm$ parents of IceCube
neutrinos decay into $\gamma$-rays, which are only observed indirectly
after propagation in the extragalactic radiation fields permeating the
universe. These $\gamma$-rays initiate inverse Compton cascades that
degrade their energy below 1~TeV. The relative magnitudes of the
diffuse $\gamma$-ray flux detected by
\emph{Fermi} LAT~\cite{Ackermann:2014usa} can then be used to constrain the
spectral index, assuming the $\gamma$-rays produced by the $\pi^0$'s
accompanying the $\pi^\pm$ escape the source. Figure~\ref{fig:fermi+icecube+dm} shows
that only a relatively hard injection spectrum is consistent with the
data.  Indeed, if IceCube neutrinos are produced through $pp$
collisions in optically thin extragalactic sources, the $\gamma$-rays
expected to accompany the neutrinos saturate the \emph{Fermi} LAT flux for
$\alpha \approx 2.2$~\cite{Murase:2013rfa}.  The overall isotropy of the
observed arrival directions and the fact that a PeV event arrives from
outside the Galactic plane disfavor a Galactic origin. Moreover,
for the Galactic hypothesis one must consider another important
caveat, namely that the expected photon flux in the PeV
range~\cite{Anchordoqui:2013qsi} has been elusive~\cite{Ahlers:2013xia}.

The difficulties so far encountered in modeling the production of
IceCube neutrinos in astrophysical sources fueled the interest in
particle physics inspired models. By far the most popular model in
this category is the decay of a heavy massive ($\sim$ few PeV) relic
that constitute (part of) the cold dark matter (CDM) in the
universe~\cite{Feldstein:2013kka}.  The lack of events in the vicinity
of the Glashow resonance, implies the spectrum should decrease
significantly at the energy of a few PeV. Spectra from dark matter
decays always exhibit a sharp cutoff determined by the particle
mass. Furthermore, the 3 highest energy events appear to have
identical energies, up to experimental uncertainties. A line in the
neutrino spectrum would be a smoking gun signature for dark matter. If
the heavy relic also decays into quarks and charged leptons, the mono-energetic neutrino
line may be accompanied by a continuous spectrum of lower-energy
neutrinos, which can explain both the PeV events and some of the
sub-PeV events. All of these considerations appear to be in agreement with the
data~\cite{Esmaili:2013gha}. Even much heavier relic particles, with
masses well above a PeV, can generate the required neutrino spectrum
from their decays if their lifetime is much shorter than the present
age of the universe~\cite{Ema:2013nda}. The spectrum of neutrinos is
modified by a combination of redshift and interactions with the
background neutrinos, and the observed spectrum can have a cutoff just
above 1~PeV for a broad range of the relic particle masses. In this Letter we
will reexamine the idea of a dark matter origin for IceCube events.

{\color{rossoCP3} {\bf $\bm{H_0}$ {\bf
    \emph{Measurements---}}}} Another, seemingly different, but perhaps closely related subject is
the emerging tension between direct astronomical measurements at low
redshift and cosmological parameters deduced from temperature
fluctuations in the cosmic microwave background (CMB). Strictly
speaking, the TT, TE, EE spectra recorded by the \emph{Planck} spacecraft
when combined with polarization maps (lowP) describe the standard
spatially-flat 6-parameter $\Lambda$CDM model $\{\Omega_b h^2,\,
\Omega_{\rm CDM} h^2,\, \Theta_s,\, \tau,\, n_s,\, A_s\}$ with high
precision: {\it (i)}~baryon density, $\Omega_b h^2= 0.02225 \pm 0.00016$;
{\it (ii)}~CDM density, $\Omega_{\rm CDM} h^2 = 0.1198 \pm 0.0015$;
{\it (iii)}~angular size of the sound horizon at recombination,
$\Theta_s = (1.04077 \pm 0.00032) \times 10^{-2}$; {\it (iv)}~Thomson
scattering optical depth due to reionization, $\tau = 0.079 \pm
0.017$; {\it (v)}~scalar spectral index, $n_s = 0.9645 \pm 0.0049$;
{\it (vi)}~power spectrum amplitude of adiabatic scalar perturbations,
$\ln (10^{10}\, A_s) = 3.094 \pm 0.034$~\cite{Ade:2015xua}.  \emph{Planck}
data also constrain the Hubble constant $h = 0.6727 \pm 0.0066$,
the dark energy density $\Omega_\Lambda = 0.6844 \pm
0.0091$, the amplitude of initial density perturbations $\sigma_8 =
0.831 \pm 0.013$,
and the mass density parameter $\Omega_m = 0.3156 \pm 0.0091$.\footnote{Throughout we adopt the usual convention of writing the
  Hubble constant at the present day as $H_0 = 100 \ h~{\rm km} \ {\rm
    s}^{-1} \ {\rm Mpc}^{-1}$.}  Unexpectedly, the $H_0$ inference
from \emph{Planck} observations deviates by more than $2.5\sigma$ from the
previous interpretation of the Hubble Space Telescope (HST) data
(based on over 600 cepheids in host galaxies and 8 samples of SNe Ia)
which leads to $h = 0.738 \pm 0.024$, including both statistical and
systematic uncertainties~\cite{Riess:2011yx}. A separate study by the Carnegie Hubble
program using  mid-infrared calibration of the cepheid distance scale
based on data from NASA's Spitzer Space Telescope yields $h = 0.743 \pm 0.021$~\cite{Freedman:2012ny}.  Besides, the
interpretation of gravitational lensing time delay measurements of the
system RXJ1131-1231 points to $h = 0.787^{+ 0.043}_{-0.045}$~\cite{Suyu:2012aa}.

The tension between the CMB based determination of the Hubble constant
and the $h$ value inferred from direct low redshift measurements is
intriguing and deserves further attention. On the one hand, the
underlying source of discrepancy could be some systematic uncertainty
in the calibration~\cite{Rigault:2014kaa}.  On the other hand, it
could trace a deficiency of the concordance model of cosmology. In the
spirit of~\cite{Dienes:2011ja}, it has been recently conjectured that
\emph{Planck}-inspired $\Lambda$CDM paradigm can be reconciled with
HST measurements if a subdominant fraction $f_X$ of CDM is unstable
and decays rather quickly with respect to the present Hubble
time~\cite{Berezhiani:2015yta}. The width of the unstable component
$\Gamma_X$ is an independent parameter of the
model. By forcing the $X$ particles to decay after recombination
$\Gamma_X$ is bounded from above. Moreover, the $X$ is assumed to
decay (dominantly) into invisible massless particles of the hidden
sector and hence does not produce too many photons. A joint fit to
\emph{Planck}, supernova, and HST data reveals that the base
$\Lambda$CDM model, with $\Gamma_X = 0$, is outside the $2\sigma$
likelihood contours in the ($\Gamma_X,f_X$)
plane~\cite{Berezhiani:2015yta,Chudaykin:2016yfk}. The data instead
favor $0.03 \alt f_X \alt 0.12$. The mean value and $1\sigma$ error
derived from a maximum likelihood analysis are $h = 0.716 \pm
0.020$~\cite{Berezhiani:2015yta}. Interestingly, within the same
parameter range the model could also alleviate the emerging tension
with the cluster data. (See, however, \cite{Angrick:2015gta}.) For
example, for $f_X \simeq 0.10$ and $\Gamma_X \simeq 2000~{\rm km} \,
{\rm s^{-1}}\, {\rm Mpc}^{-1}$ the corresponding values of $\Omega_m
\simeq 0.25$ and $\sigma_8 \simeq 0.80$~\cite{Berezhiani:2015yta} are
marginally consistent with the $2\sigma$ allowed contours by the
\emph{Planck} cluster mass scale~\cite{Ade:2015fva} and the extended
ROSAT-ESO Flux Limited X-ray Galaxy Cluster Survey (REFLEX
II)~\cite{Bohringer:2014ooa}. For smaller values of $f_X$ and or
$\Gamma_X$ the values of $\Omega_m$ and $\sigma_8$ move closer to
the base $\Lambda$CDM model~\cite{Berezhiani:2015yta}.  Next, in line
with our stated plan, we take $f_X \simeq 0.05$ and $\Gamma_X 
\simeq 5000~{\rm km} \, {\rm s^{-1}}\, {\rm Mpc}^{-1}$ (favored by
lensing constraints~\cite{Chudaykin:2016yfk}) as benchmarks and
investigate what would be the CDM fraction required to decay into the
visible sector to accommodate IceCube observations.

{\color{rossoCP3} {\bf  \emph{Bump-Hunting---}}} The two main parameters characterizing the $X$ particle are its lifetime $\tau_X \simeq 6
\times 10^{15}~{\rm s}$ and its mass $m_X$, which is a free
parameter. We assume the neutrino produced via $X$ decay is
mono-energetic, with energy $\varepsilon_\nu = m_X/2$. The neutrino energy distribution
from $X$ decay is given by $dN_\nu/dE_\nu = N_\nu \ \delta (E_\nu -
\varepsilon_\nu),$ where $N_\nu$ is the neutrino multiplicity. We
further assume the dominant decay mode into the visible sector,
contributing to neutrino production, is $X \to \nu \bar \nu$ and so
$N_\nu = 2$.

The evolution of the number
density of neutrinos $n_\nu (E_\nu,z)$ produced at 
cosmological distances in the decay of $X$ particles is governed by the Boltzmann continuity equation,
\begin{equation}
\frac{\partial [n_\nu/(1+z)^3]}{\partial t} = \frac{\partial [HE_\nu
  n_\nu/(1+z)^3]}{\partial E_\nu} +\mathcal{Q}_\nu \,,
\label{evolution-eq}
\end{equation}
together with the Friedman-Lema\^{\i}tre equations describing the
cosmic expansion rate $H(z)$ as a function of the redshift $z$. This
is given by $H^2 (z) = H^2_0\,[\Omega_m (1 + z)^3 +
\Omega_{\Lambda}]$. The time-dependence of the red-shift can be
expressed via $dz = -d t\,(1+z)H$.  We have found that for the
considerations in the present work neutrino interactions on the cosmic
neutrino background can be safely neglected~\cite{Weiler:1982qy}. In
(\ref{evolution-eq}),
\begin{equation}
{\cal Q}_\nu (E_\nu,t) =  \frac{n_X(t)}{\tau_X} \, {\cal B}_{X \to
\nu \bar \nu} \ \frac{d
  N_\nu}{dE_\nu} \,,
\end{equation}  
is the source term, $n_X (t) = Y_X \ s(t) \ e^{-t/\tau_X}$ is the
number density of $X$, ${\cal B}_{X \to \nu \bar \nu}$ is the neutrino
branching fraction, $s(t)$ is the entropy density, and
\begin{equation}
  Y_X = 3.6 \times 10^{-9} \frac{\Omega_X h^2}{m_X/{\rm GeV}} 
\label{yield}
\end{equation}
 is the comoving number
density at the CMB epoch.  By solving (\ref{evolution-eq}) we obtain
the (all flavor) neutrino flux at present epoch $t_0$,
\begin{eqnarray}\label{flux}
\Phi (E_\nu) & = & \frac{c}{4\pi} n_\nu (E_\nu,0) \\
& = &\frac{c}{4\pi} \frac{N_\nu Y_X s(t_0)}{\tau_X E_\nu}  
{\cal B}_{X \to \nu \bar \nu}
\left. \frac{e^{-t_*/\tau_X}}{H(t_*)} \right|_{1+z(t_*) =
 \varepsilon_\nu/E_\nu }  , \nonumber
\end{eqnarray}
with $s(t_0) \simeq 2.9 \times 10^3~{\rm cm^{-3}}$. 

Maximization of (\ref{flux}) yields the energy relation 
for the peak in the spectrum,
\begin{equation}
E_\nu^{\rm peak} \simeq  \frac{1}{2} \ \frac{m_X/2}{1+
z(\tau_X)} \,,
\end{equation}
which sets the mass of the $X$. Since $z(\tau_X) \simeq 18$ to
accommodate the PeV peak in IceCube's neutrino spectrum we take $m_X
\simeq 76~{\rm PeV}$. Now, from (\ref{yield}) we obtain
$Y_X \approx 2.8 \times 10^{-19}$. Finally, we normalize the cosmic
neutrino flux per flavor using (\ref{icecubeflux}). The intensity of
the mono-energetic signal at the peak is taken as 60\% of the flux
reported by the IceCube Collaboration, yielding a neutrino branching
fraction ${\cal B}_{X \to \nu \bar \nu} \sim 3 \times 10^{-9}$ into
all three flavors. The width, an output of the Boltzmann equation, is
shown in Fig.~\ref{fig:fermi+icecube+dm}. It is evident that the
mono-energetic neutrino spectrum is in good agreement with the
data. In particular, the flux suppression at the Glashow resonance,
$\Phi (E_\nu^{\rm res})/\Phi(E_\nu^{\rm peak}) \simeq 0.011$, is
consistent with data at $1\sigma$.

The model is fully predictive and can also be confronted with
\emph{Fermi} LAT data. It is reasonable to assume that ${\cal B}_{X \to
  e^+ e^-} \approx {\cal B}_{X \to \nu_e \bar \nu_e} \approx {\cal
  B}_{X \to u \bar u} \approx {\cal B}_{X \to d \bar d}$. About 1/3 of
the energy deposited into either $u\bar u$ or $d \bar d$ is channeled
into $\gamma$-rays via $\pi^0$ decay and about 1/6 of the energy is
channeled into electrons and positrons.  As previously noted, the
$\gamma$-rays, electrons, and positrons trigger an electromagnetic cascade on
the CMB, which develops via repeated $e^+ e^-$ pair production and
inverse Compton scattering.  As a result of this cascade the energy gets
recycled yielding a pile up of
$\gamma$-rays at ${\rm GeV} \alt E_\gamma \alt {\rm TeV}$, just below
the threshold for further pair production on the diffuse optical
backgrounds.  We have seen that under very reasonable assumptions the
energy deposited into neutrinos is comparable to the energy deposited
into the electromagnetic cascade. Therefore, the neutrino energy
density at the present epoch,
\begin{equation}
\omega_\nu = \int E_\nu \ n_\nu (E_\nu, 0) \ dE_\nu = 2.6 \times
10^{-11}~{\rm eV} \ {\rm cm}^{-3},
\end{equation}
provides reliable estimate of the cascade energy density ($\omega_{\rm
  cas} \sim \omega_\nu$), which is bounded by \emph{Fermi} LAT data to
not exceed $\omega_{\rm cas}^{\rm max} \sim 5.8 \times 10^{-7}~{\rm
  eV} \, {\rm cm^{-3}}$~\cite{Berezinsky:2010xa}. We conclude that the
$\gamma$-ray flux associated with the neutrino line is found to be
about 4 orders of magnitude smaller than the observed flux in the
\emph{Fermi} LAT region.

We now turn to discuss the possibility of distinguishing the neutrino
line from an unbroken power-law spectrum without the neutrino line,
with future IceCube data.  The value of the spectral index is
determined by the ``low energy'' events. Following the best IceCube
fit we adopt a spectrum $\propto E_\nu^{-2.46}$.  We assume that the
IceCube events below 1~PeV have an astrophysical origin. Indeed, the
steep spectrum $\propto E_\nu^{-2.46}$ may suggest we are witnessing
the cutoff of TeV neutrino sources running out of power. Using the
IceCube aperture for the high-energy starting event (HESE)
analysis~\cite{Aartsen:2013bka} we compute the event rate per year
above 1~PeV for both the neutrino flux given in (\ref{icecubeflux})
and that of (\ref{flux}).  The results are given in
Table~\ref{table1}. As expected, the predictions from $X$ decay are in
good agreement with existing data. Because of the smeared
energy-dependence of muon tracks, in what follows we will only
consider cascades and double bang topologies initiated by charged
current interactions of electron and tau neutrinos, as well as all
neutral current interactions processes. We identify the events coming
from the power law spectrum ${\cal N}_{\rm B}$ with background and
adopt the standard bump-hunting method to establish the statistical
significance of the mono-energetic signal.  To remain conservative we
define the noise $\equiv \sqrt{{\cal N}_{\rm B} + {\cal N}_{\rm S}}$,
where ${\cal N}_{\rm S}$ is the number of signal events. In 10 years of
operation the total detection significance,
\begin{equation}
S_{\rm det} = \frac{{\cal N}_{\rm S}}{\sqrt{{\cal N}_{\rm B} + {\cal
      N}_{\rm S}}} \,,
\end{equation}
would allow distinguishing the neutrino line from a statistical
fluctuation of a power law spectrum $\propto E_\nu^{-2.46}$ at the
$3\sigma$ level. Note that the shape of the distribution with energy
conveys additional information allowing one to distinguish the line
signal from fluctuations of a power-law background.  The proposed
\emph{IceCube-Gen2} extension plans to increase the effective volume
of IceCube by about a factor of 10~\cite{Aartsen:2014njl}. This
facility will not only increase the HESE sensitivity but also improve
the energy resolution for muon tracks.  In a few years of operation
\emph{IceCube-Gen2} will collect enough statistics to elucidate the
dark matter--neutrino connection with $S_{\rm det} > 5\sigma$.

\begin{table}[t]
\vspace*{-0.1in}
\caption{Event rates (${\rm yr^{-1}}$) at IceCube for $E_\nu^{\rm min}
  = 1~{\rm PeV}$.}
\label{table1}
\begin{tabular}{c|@{}ccc|@{}ccc}
\hline
\hline
~~$E_\nu^{\rm max}/{\rm PeV}$~~ & \multicolumn{3}{@{}c|}{~~spectrum
  $\propto E_\nu^{-2.46}$~~} & \multicolumn{3}{@{}c}{~~$X$ decay spectrum~~} \\
\hline
 & ~~$\nu_e$~~ & ~~$\nu_\mu$~~ & ~~$\nu_\tau$~~ & ~~$\nu_e$~~ &
 ~~$\nu_\mu$~~ & ~~$\nu_\tau$~~ \\
\cline{2-4} \cline{5-7}
2 &~~0.20~~&~~0.18~~&~~0.20~~&~~0.25~~&~~0.23~~&~~0.25~~\\
3 &~~0.27~~&~~0.24~~&~~0.27~~&~~0.46~~&~~0.41~~&~~0.46~~\\
4 &~~0.31~~&~~0.27~~&~~0.30~~&~~0.58~~&~~0.51~~&~~0.56~~\\
5 &~~0.34~~&~~0.29~~&~~0.32~~&~~0.66~~&~~0.56~~&~~0.62~~\\
\hline
\hline
\end{tabular}
\end{table}

We end with an observation: IceCube data can also be fitted by a
neutrino line peaking at $E_\nu \sim 20~{\rm TeV}$ superimposed over
a power law spectrum \mbox{($\propto E_\nu^{-2}$)} of astrophysical
neutrinos~\cite{Aisati:2015vma}.  By duplicating our discussion for
$m_X \sim 1~{\rm PeV}$ it is straightforward to see that the model can
also accommodate this neutrino line.

{\color{rossoCP3} {\bf \emph{Conclusions---}}} We have shown that the
PeV flux of extraterrestrial neutrinos recently reported by the
IceCube Collaboration can originate through the decay of heavy dark
matter particles with a mass $\simeq 76~{\rm PeV}$ and a
lifetime $\simeq 6 \times 10^{15}~{\rm s}$.  On a separate track, the
tension between \emph{Planck} data and low redshift astronomical
measurements can be resolved if about 5\% of the CDM component at CMB
epoch is unstable. Assuming that such a fraction of quasi-stable relics is
responsible for the IceCube flux we determined the neutrino branching
fraction, ${\cal B}_{X \to \nu \bar \nu} \sim 3 \times 10^{-9}$.  The
model has no free parameters and will be tested by future IceCube
data. Indeed 10 years of data taking will be required to distinguish
the neutrino line from an unbroken power-law spectrum at the $3\sigma$
level. The upgraded \emph{IceCube-Gen2} will collect enough statistics
to elucidate the dark matter--neutrino connection at the $5\sigma$
discovery level in a few years of operation.

{\color{rossoCP3} {\bf \emph{Acknowledgments---}}} LAA is supported by
 U.S. National Science Foundation (NSF) CAREER Award PHY1053663 and
 by the National Aeronautics and Space Administration (NASA) Grant
 No. NNX13AH52G; he thanks the Center for Cosmology and Particle
  Physics at New York University for its hospitality. VB is supported
  by the U. S. Department of Energy (DoE) Grant No. DE- FG-02-
  95ER40896. HG is supported by NSF Grant No. PHY-1314774.  XH is
  supported by the MOST Grant 103-2811-M-003-024. DM is supported by
  DoE Grant No. DE-SC0010504. TJW is supported by DoE Grant
  No. DE-SC0011981.

\end{document}